# WIRE COMPENSATION OF PARASITIC CROSSINGS IN DAΦNE

C. Milardi, D. Alesini, M.A. Preger, P. Raimondi, M. Zobov, LNF-INFN, Frascati, Italy
D. Shatilov, BINP, Novosibirsk, Russia

## Abstract

Long-range beam-beam interactions (parasitic crossings) are one of the main luminosity performance limitations for the Frascati $e^+e^-$ Φ-factory DAΦNE. In particular, the parasitic crossings lead to a substantial lifetime reduction of both beams in collision. This puts a limit on the maximum storable current and, as a consequence, on the achievable peak and integrated luminosity. In order to alleviate the problem numerical and experimental studies of the parasitic crossings compensation with current-carrying wires have been performed at DAΦNE. Two such wires have been installed at both ends of the KLOE interaction region. Switching on the wires in agreement with the numerical predictions, improvement in the lifetime of the "weak" beam (positrons) has been obtained at the maximum current of the "strong" one (electrons) without luminosity loss. In this paper we describe the parasitic crossings effects in DAΦNE, summarize the results of numerical simulations on their compensation with the wires and discuss the experimental measurements and observations.

## INTRODUCTION

The Frascati Φ-factory DAΦNE is an $e^+e^-$ collider operating at the energy of Φ-resonance (1.02 GeV c.m.) [1]. Its best peak luminosity reached so far is $1.5 \times 10^{32}$ cm$^{-2}$s$^{-1}$ with a maximum daily integrated luminosity of about 10 pb$^{-1}$ [2].

In order to obtain such a high luminosity at low energy high current bunched beams are stored in each collider ring. Usually, the number of adjacent filled buckets is in the range 109÷111 out of 120 available. A short gap is needed for ion clearing. It's worth reminding that in DAΦNE the bunch separation of 2.7 ns is the shortest among all existing colliders and particle factories.

In order to minimize the effect of parasitic crossings (PC) of the closely spaced bunches the beams collide under a crossing angle in the range 10÷20 mrad. However, in spite of the crossing angle, the long-range beam-beam interactions (LRBB) remain one of the most severe limitations to the performance of DAΦNE in terms of luminosity. In fact LRBB interaction leads to a substantial lifetime reduction of both beams, limiting the maximum storable currents and, as a consequence, the maximum achievable peak and integrated luminosity. The latter is strongly influenced by the beam lifetime because in the topping up regime a fraction of the integrated luminosity is lost during the time necessary to switch the injection system from from one beam to the other. The problem of the LRBB interactions is expected to be even more pronounced in future operation for the FINUDA experiment [3], because the crossing angle is intrinsically lower. The PC effects are foreseen to be also very important for the DAΦNE upgrade options [4], where the number of colliding bunches will be increased by means of higher frequency RF cavities making the bunch separation even shorter than in the present situation.

Looking for a compensation scheme to reduce the impact of LRBB interactions we decided to install two windings (wires) at both ends of the KLOE interaction region (IR). This approach revises an idea originally proposed by J.P.Koutchouk [5] for LHC, and recently tested during single beam operation on SPS [6, 7]. Simulations using LHC compensation devices also predicted a relevant dynamic aperture enlargement for the Tevatron collider [8].

At DAΦNE we have performed an experimental study of the LRBB compensation using purposely built wires obtaining encouraging results, and testing for the very first time the wire compensation scheme in collision.

In this paper we describe our experience and the experimental results.

## PARASITIC CROSSINGS IN DAΦNE

DAΦNE consists of two independent rings sharing two interaction regions: IR1 and IR2. The KLOE detector is installed in IR1. While delivering luminosity to KLOE bunches collide with a crossing angle of 14.5 mrad, and are vertically separated in IR2 by a distance larger than 200 $\sigma_y$. For this reason, in the following considerations only LRBB interactions in IR1 will be taken into account.

The bunches experience 24 PCs in IR1, 12 before and 12 after the main interaction point (IP), until they are separated by splitter magnets into two different rings.

Table 1 summarizes the main parameters for some PCs: relative position, beta functions, phase advances with respect to the IP and transverse separation in terms of $\sigma_{x,y}$.

Table 1: Parameters for the Pcs, one every four, in IR1.

| PC order | $Z-Z_{IP}$ [m] | $\beta_x$ [m] | $\beta_y$ [m] | $\mu_x-\mu_{IP}$ | X [$\sigma_x$] | Y [$\sigma_y$] |
|---|---|---|---|---|---|---|
| BB12L | -4.884 | 8.599 | 1.210 | 0.167230 | 26.9050 | 26.238 |
| BB8L | -3.256 | 10.177 | 6.710 | 0.140340 | 22.8540 | 159.05 |
| BB4L | -1.628 | 9.819 | 19.416 | 0.115570 | 19.9720 | 63.176 |
| BB1L | -0.407 | 1.639 | 9.426 | 0.038993 | 7.5209 | 3.5649 |
| IP1 | 0.000 | 1.709 | 0.018 | 0.000000 | 0.0000 | 0.0000 |
| BB1S | 0.407 | 1.966 | 9.381 | 0.035538 | -6.8666 | 3.5734 |
| BB4S | 1.628 | 14.447 | 19.404 | 0.092140 | -16.4650 | 63.196 |
| BB8S | 3.256 | 15.194 | 6.823 | 0.108810 | -18.7050 | 157.74 |
| BB12S | 4.884 | 12.647 | 1.281 | 0.126920 | -22.1880 | 25.505 |

There are indications that the LRBB interactions affect the beam dynamics in DAΦNE: in fact, the PCs induce orbit distortion that can be satisfactory reproduced by the machine model, based on the MAD [9] code, when the PCs are taken into account. MAD predictions agree with the orbit distortion obtained from the beam-beam simulation code Lifetrack [10], see Fig. 1

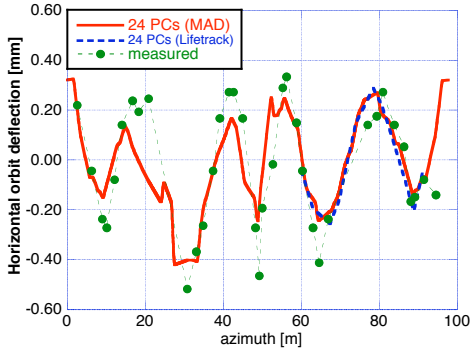

Figure 1: measured and computed orbit deflection due to 24 BBLR interactions for the positron bunch colliding against 10 mA electron bunches.

Moreover, the lifetime of each beam starts decreasing during injection of the opposite beam and remains low soon after injection. Typically, in collision, the electron beam current reaches 1.8÷2.2 A, while the maximum positron beams is 1.3÷1.4 A. Exceeding these values the lifetime of the beams drops down to 700÷800 sec. Presently this behaviour is considered one of the main limitation of the collider performance.

## NUMERICAL SIMULATIONS

The "weak-strong" tracking code LIFETRAC was used to simulate the equilibrium distribution of the positron ('weak') beam. The main sources of long beam tails are the 2 PCs nearest to the Main IP, but the other PCs also make some contribution, so we account for all of them. The wires were simulated as additional PCs with variable current ("wire-PC"), so that no special tracking algorithm for wires was used. This approach is justified by the rather large values of the $\beta_{x,y}$ functions at the wire locations (16.5 and 4 m respectively), much larger than both the bunch and wire length. This allows simulating the interaction with the wire as a single kick, neglecting the effect of movement of the "strong" bunch: the longitudinal coordinate of collision point for the real PC depends on the particle's longitudinal coordinate, while the wire are fixed, but for large betas a shift of few millimeters gives actually no effect. On the other hand, the betas are small enough to have a large separation in units of the transverse beam size (≈20), so the actual "shape of wire" (i.e. density distribution inside the wire-PC) does not matter: it works like a simple 1/r lens. Some simulation results are shown in Fig. 2. The beam current was chosen to be large enough to have long beam tails due to PCs (a), then the wires were switched on and the tails reduced (b). When the wires are powered with wrong polarity, the tails blow-up becomes even stronger (c).

As a matter of fact, the PCs compensation with a single wire on each side of the interaction region is not perfect since distances between the beams at PC locations are different in terms of the horizontal sigma and phase advances between PCs and wires are not completely compensated (see Table 1). Indeed, the numerical simulations did not show improvements in luminosity. However, the positive effect of tails reduction and corresponding lifetime increase is very important, because it leads to a larger integrated luminosity.

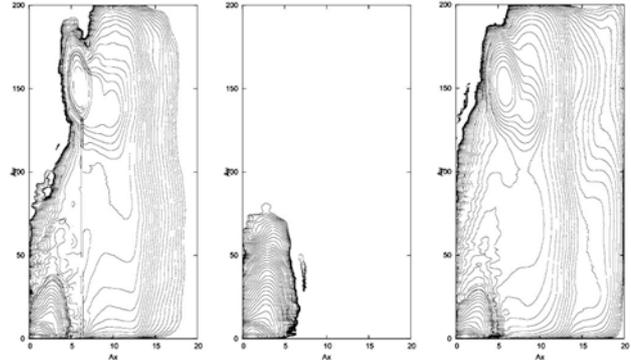

Figure 2: Particle equilibrium density in the normalized transverse phase space, starting from left: wires off (a), wires on (b) and wires powered with wrong polarity (c).

## WIRE DESIGN AND INSTALLATION

The wires have been built and installed in IR1 in November 2005. Each device is made of two coils of rectangular shape, 20 windings each, installed symmetrically with respect to the horizontal plane, see Fig. 3. Our device differs from the LHC one for several aspects: our wires are installed outside the vacuum chamber exploiting a short section in IR1, just before the splitters, where the vacuum pipes are separated to host Lambertson type correctors not essential for operation and therefore removed. The wires carry a tunable DC current, and produce a stationary magnetic field with a shape similar to the one created by the opposite beam.

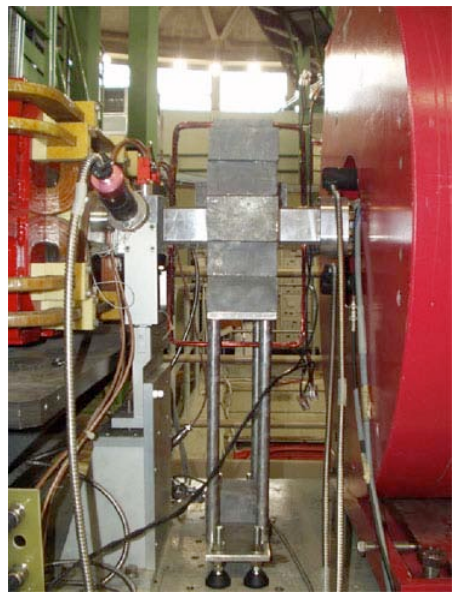

Figure 3: The wires installed at one end of IR1.

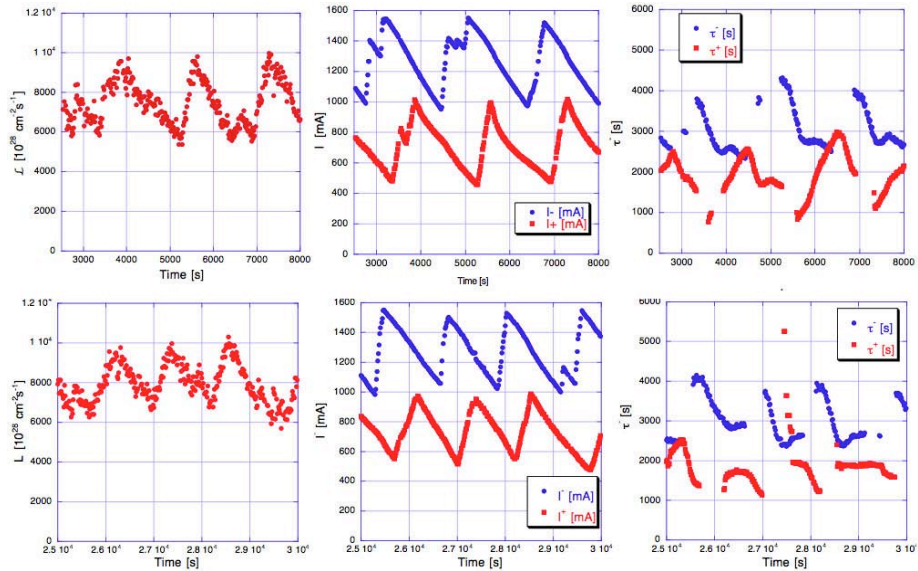

Figure 4: Luminosity, colliding currents and lifetime as a function of time : wires on (upper frames) and wires off (lower frames).

## EXPERIMENTAL RESULTS

A systematic study of the wires in collision has been undertaken during the machine shifts in March 2006.

The wires were powered at 3.6 A to compensate as much as possible the beam lifetime of the positron beam that, due to the limited maximum achievable current, can be considered as the 'weak' one.

It has been experimentally verified that the residual orbit distortions with maximum deviations of +0.4 -0.5 mm due to the PCs is very well corrected with wire currents of ≈1 A. This is a proof that the wires behave as correctors "in phase" with the PCs. It has been also measured that the wires introduce some betatron tunes shifts which should be corrected.

The residual orbit distortion due to the wires at 3.6 A was corrected by the ordinary dipole correctors, while the tune shifts were compensated by means of the quadrupoles in a dispersion free straight section.

Several luminosity runs have been compared by switching the wires on and off to study their impact on the collisions. In the following the two most relevant sets are presented taking into account 2 hours long runs. The results in Fig. 4 show some clear evidences. Switching on and off the wires we obtain the same luminosity while colliding the same beam currents. The positron lifetime is on average higher when wires are on, while the electron one is almost unaffected. The beam blow-up occurring from time to time at the end of beam injection, corresponding to a sharp increase in the beam lifetime, almost disappear.

A further aspect becomes evident when comparing, on the same plot, the positron current and lifetime with and without wires, see Fig. 5. The positron current starts from the same value, then, in the case of wires off the current decreases with a higher derivative than in the case with wires on. The lifetime exhibits a consistent behavior, being higher in the case when wires are on. In this way it is possible to keep the same integrated luminosity injecting the beam two times only instead of three in the same time integral, or to increase the integrated luminosity by the same factor keeping the same injection rate.

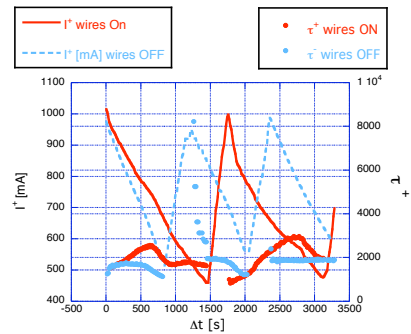

Figure 5: Positron current and lifetime as a function of time: wires on (red) and wires off (cyan).

## CONCLUSIONS AND ACKNOWLEDGMENTS

The wires installed on DAΦNE proved to be effective in reducing the impact of BBLR interactions and improving the lifetime of the positron beam. We are indebted to G. Sensolini, R. Zarlenga and F. Iungo for the technical realization of the device.